\documentclass[aps,prb,twocolumn,groupedaddress,floatfix,showpacs,showkeys]{revtex4}
\usepackage{amssymb}
\usepackage{amsmath}
\usepackage{amsfonts}
\usepackage[dvips]{graphicx}
\usepackage[T1]{fontenc}
\graphicspath{{/Ciencia/Articles_Oscar/JNN_2006/},{/Ciencia/Articles_Oscar/JNN_2006/Article/}}

\begin{document}


\title{Exchange bias phenomenology and models of core/shell nanoparticles}


\author{\`Oscar Iglesias}
\email{oscar@ffn.ub.es}
\homepage{http://www.ffn.ub.es/oscar}
\author{Am\'{\i}lcar Labarta}
\author{Xavier Batlle}
\affiliation{Departament de F\'{\i}sica Fonamental and Institut de Nanoci\`encia i Nanotecnologia (IN2UB), Universitat de Barcelona, Diagonal 647, 08028 Barcelona, Spain}


\date{\today}

\begin{abstract}
Some of the main experimental observations related to the occurrence of exchange bias in magnetic systems are reviewed, focusing the attention on the peculiar phenomenology associated to nanoparticles with core/shell structure as compared to thin film bilayers. The main open questions posed by the experimental observations are presented and contrasted to existing theories and models for exchange bias formulated up to date. We also present results of simulations based on a simple model of a core/shell nanoparticle in which the values of microscopic parameters such as anisotropy and exchange constants can be tuned in the core, shell and at the interfacial regions, offering new insight on the microscopic origin of the experimental phenomenology. A detailed study of the of the magnetic order of the interfacial spins shows compelling evidence that most of the experimentally observed effects can be qualitatively accounted within the context of this model and allows also to quantify the magnitude of the loop shifts with striking agreement with the macroscopic observed values.
\end{abstract}

\pacs{75.60.-d,05.10.Ln,75.50.Tt,75.60.Jk}
\keywords{magnetic nanoparticles; core/shell nanoparticles; exchange bias; Monte Carlo simulation}

\maketitle


\section{Introduction}

Magnetic fine particles have attracted a constant interest among the scientific community during the last decades because of their increasing number of applications \cite{Dormann_acp97}. The demand for miniaturization and the availability of new synthesis and measurement techniques \cite{Martin_jmmm03} have allowed to prepare nanostructructured materials with different dimesionalities on the submicron range. This has open the field of nanomagnetism to a handful of new opportunies \cite{Bader_rmp06} that exploit new magnetic, optical and electrical properties that emerge when reducing the size of the particles to the nanoscale \cite{Prinz_science98,Moriarty_rpp01,Skomski_jpcm03}, of interest in wide areas of science ranging from magnetic recording and quantum computing \cite{DiVicenzo_jmmm99} to Earth sciences \cite{Rancourt_rmg01} and biomedicine \cite{Pankhurst_jpd03,Berry_jpd03,Tartaj_jpd03,Tartaj_jmmm05,Tartaj_cn06}.

Due to their reduced dimensions, nanoparticles display peculiar magnetic and transport properties \cite{Batlle_jpd02} that are not present in the bulk materials as a consequence of the interplay between intrinsic properties arising from finite-size effects and collective effects due to different kinds of interparticle interactions \cite{Dormann_jmmm99}.
A direct consequence of the finite size of the particles is superparamagnetism, which is a drawback for magnetic recording applications because it causes thermal destabilization of the recording units. However, superparamagnetic (SP) response is desirable for most of biomedical applications.
Another effect influencing the magnetic response of the nanoparticles is the reduction of the magnetic net moment as compared to bulk. This is due to the competition between the different magnetic ordering at the particle core and its surface, which has a higher degree of disorder due to the broken symmetry, roughness and different stoichiometry from the bulk material. 
Particle surfaces are usually exposed to environment and are, therefore, easily oxydized, resulting in core/shell structures that can be otherwise produced by controlled chemical synthesis \cite{Willard_imr04,Darling_jmc05} in a variety of morphologies and compositions. Magnetic core/shell nanoparticles with functionalyzed shells and coatings are also necessary in biomedicine for applications in targeted delivery and diagnostics 
\cite{Krishnan_jmc06}. 

An attractive composition results from the combination of a ferromagnetic (FM) core surrounded by an antiferromagnetic (AFM) shell (usually an oxide) coupled by the exchange interaction at the interface between them. Interesting proximity effects result from the structural modification and competition of different magnetic orderings at the FM/AFM interface. 
In particular, the exchange coupling at a FM/AFM interface may induce unidirectional anisotropy in the FM below the Ne\'el temperature of the AFM, causing a shift in the hysteresis loop, a phenomenon known as exchange bias (EB). For EB to occur, the Curie temperature $T_C$ of the FM has to be greater than $T_N$ and the system has to be cooled from a starting temperatute in between in the presence of an applied field $H_\mathrm{FC}$. Moreover, the anisotropy of the AFM has to be high enough so that its spins remain fixed during the hysteresis loop.
Although the first observations of this phenomenon, dating back five decades ago \cite{Meiklejohn_pr56,Meiklejohn_pr57}, were reported on oxidized nanoparticles, most of the subsequent studies have focused on layered FM/AFM structures \cite{Nogues_jmmm99,Berkowitz_jmmm99} because of their application in advanced magnetic devices \cite{Prinz_science98,Bobo_jpcm04}.
However, in recent years, the study of EB in nanoparticles and nanostructures  has gained renewed interest \cite{Nogues_PhysRep05} since it has been shown that control of the core/shell interactions or of the exchange coupling between the particle surface and the embedding matrix can be a way to beat the SP limit \cite{Skumryev_nature03,Nogues_prl06}.  

Both nanoparticles and layered systems display common phenomenolgy although, in the later case, a wider range of experimenal techniques have been used which have provided deeper knowledge on the microscopic mechanisms that are at the basis of the EB effect.
Thus, knowledge of the magnetic structure at the interface has become a subject of primary interest in understanding EB. At difference from layered systems, the interface of core/shell nanoparticles naturally incorporates roughness and non-compensation of the  magnetization, two of the main ingredients for which different assumptions are adopted by the existing models for EB in films \cite{Stamps_jpd00,Kiwi_jmmm01}. 
However, the interpretation of the results may be hindered by collective effects and interactions with the embedding matrix since, up to date, no EB experiment has been conducted on a single particle, which would allow to confront the results with the existing models.

Most of the theoretical framework for the explanation of EB is based in macroscopic or phenomenological models for layered systems, adapted to the particular structure and composition of specific combinations of materials. Guided by simplicity and reproducibility of experimental results, simplifying assumptions about the magnetic order in the FM and AFM layers are often assumed which may not allow to understand the real microscopic origin of the EB effects. 
Moreover, and despite the similarities in both cases, the models used for EB in layered systems are not well suited for particle systems, since surface effects and the reduced dimensionality of the nanoparticles are supposed to play a role in the observation of EB.

For this purpose, computer simulations based either on Monte Carlo (MC) methods or on the micromagnetic approach \cite{Fidler_jpd00} have proved useful to gain insight into the microscopic origin of EB. These methods allow to take as inputs microscopic parameters such as exchange and anisotropy constants specific to the materials at hand and also to take into account the specific arrangement of the magnetic atoms in a lattice. As an output, macroscopically measurable quantities, such as the magnetization, can be computed without loosing valuable information about the microsocopic magnetic configurations that are at the origin of the observed phenomena.

In this article, we will review the main phenomenology associated to EB in core/shell nanoparticle systems and the main existing models to explain it. The review is organized as follows. In Sec. II, we review the main nanoparticle systems for which EB has been reported, with special emphasis in nanoparticles with core/shell structure. Next, in Sec. III, we present a summary of the different phenomenology associated to EB found experimentally for core/shell nanoparticles. In Sec. IV the results of MC simulations of a model of core/shell nanoparticle recently proposed by us \cite{Iglesias_prb05,Iglesias_phyab06}, together with other results in the literature, are presented. We end with the final conclusions and remarks in Sec. V.

\section{Core/shell nanoparticles displaying EB}

Observation of EB in nanoparticles has been reported for a wide variety of materials and morphologies which can be divided in three categories: (1) single phase ferrimagnetic or antiferromagnetic oxides, (2) nanoparticles embedded in a AFM matrices and (3) nanoparticles with core/shell structure. 

In the first group we have ferrites, manganites and antiferromagnetic particles (see tables 1 and 2 for a summary of results in Ref. \onlinecite{Nogues_PhysRep05}). 
The origin of EB in this kind of nanoparticles is not established yet, since, in this case, one cannot strictly speak of a FM coupled to an AFM material. Therefore, the observation of EB has been attributed to the freezing of a spin-glass surface layer of spins which is formed due to finite-size and surface effects \cite{Punnoose_prb01,Zysler_phaB06}. However, the fact that for this kind of particles high field irreveribilities and non-saturating hysteresis loops are commonly found, poses the question of whether minor loop effects could also be at the origin of loop shifts.

Another way to get a high density of interface coupling FM and AFM phases of different materials is by embedding FM particles in AFM hosts synthesized with different techniques, although in these systems no clear separation between core and a well defined shell can be made. We refer the reader to Refs. \onlinecite{Nogues_PhysRep05} (section 3.3) and \onlinecite{Nogues_ijn05} for recent reviews of results in this kind of systems.  

Finally, some of the largest observed EB fields have been reported for particles consisting of a FM core and an AFM (or ferrimagnetic) shell which has been grown around the core by chemical modification (usually partial oxidation) of the FM material. Among them, some particular combinations have deserved special attention as 
Co/CoO, where the EB effect was first described by Meiklejohn and Bean  \cite{Meiklejohn_pr56,Meiklejohn_pr57,Meiklejohn_jap62} and revisited some decades later by Gangopadhyay et al. \cite{Gangopadhyay_nanostr92,Gangopadhyay_jap93} and later by Peng and coworkers \cite{Peng_pssa99,Peng_apl99,Peng_prb99,Peng_prb00,Sumiyama_stam05}. More recent studies of EB phenomenology in Co/CoO nanoparticles are listed in what follows \onlinecite{Wiedwald_thesis,Wiedwald_prb03,Wiedwald_prb03,Spasova_jm04,Wiedwald_prb04,Wiedwald_phtr05,Wen_jmmm04,Zhou_apa05,Verelst_chemmat99,Skumryev_nature03,Riveiro_apl05,DeToro_prb06,Palasantzas_aem05,Gruyters_prl05,Koch_prb05,Tracy_prb05,Dobrynin_apl05,Dobrynin_prb06,Portemont_jap06,Nogues_prl06,Normile_jap06,Rohart_prb06,Tracy_prb06}.
Other core/shell particle systems having Co and other oxides have also been studied such as NiCo/NiCoO \cite{Jeyadevan_ie02}, Co/CoN \cite{Lin_nanostrmat95}, 
Co/MnO \cite{VanLierop_jmmm03}, Co$_{80}$Ni$_{20}$/oxide \cite{Luna_nano04,Vazquez_phab04} and CoPt/CoO \cite{Tomou_jap06}.
Studies of iron oxidized particles such as Fe$_3$O$_4$/FeO \cite{Das_jap02,DelBianco_jmmm05}, 
Fe/$\gamma$-Fe$_2$O$_3$ \cite{DelBianco_prb02,DelBianco_jmmm03,DelBianco_prb04,Fiorani_prb06,Zheng_jap04,Zheng_prb04,Redl_jacs04, Boubeta_prb06,Bomati_Small06} and 
Fe/Fe$_x$O \cite{Gangopadhyay_prb92,Hou_jmmm98,Peng_jap02,Baker_jmmm04,Baker_jap04,Morel_jap04,Redl_jacs04,Ceylan_jap06} have also reported a variety of effects related to EB.
Let us mention also the cases of Ni/NiO \cite{Yao_nanostr95,Yao_msea96,Sako_jvstb97,Seto_jpcb05,Parada_cm06,Liu_jmmm06,Li_jac06},  
Cr$_2$O$_3$/CrO$_2$ \cite{Zheng_apl04}, ZnFe$_2$O$_4$/CoFe$_2$O$_4$ \cite{Masala_ssc06},
FeCo/CoFeO \cite{Gangopadhyay_jap94},
and also recently FePt/MnO \cite{Kang_jacs06}, and FePt/Fe$_3$O$_4$ \cite{Zeng_nl04,Zeng_apl04}.  
There has been also recent reports of EB in unconventional morphologies such as AFM Mn (core)/ferrimagnetic (FIM) Mn$_3$O$_4$(shell) \cite{Si_apl05}, FIM CoFe$_2$O$_4$ (core)/AFM Mn (shell) \cite{Masala_jacs05} nanoparticles, Fe$_3$O$_4$/Co nanocables \cite{Kazakova_prb06,Salabas_nanol06}, and even Fe/Co oxidized particles encapsulated in a ferritin case \cite{Klem_jacs07}.    

\section{EB phenomenology}

Although the main indication of the existence of exchange bias is the observation of shifted hysteresis loops along the field axis after field cooling across the Neél temperature of the AFM $T_N$, some other macroscopic effects usually accompany the observation of loop shifts. In what follows, we will summarize the main experimental  observations related to EB peculiar to core/shell nanoparticles, comparing them with similar results observed in layered systems when possible.

\subsection {Coercivity increase}
The most usual is the increase in the coercive field $H_C$ after field cooling observed below $T_N$, which is related to the unidirectional anisotropy induced on the FM by the field cooling process. Increased coercivities should appear only when the anisotropy of the AFM component is small compared with the exchange coupling with the FM component. In this case, partial rotation of the spins of the AFM shell, which are dragged by the FM core spins during the hysteresis loop, is expected resulting in increased $H_C$.  A two times increase was found in for 13 nm Co/CoO in Ref. \cite{Wiedwald_phtr05} and also in Ref. \cite{Lee_apl06}.

\subsection {Partice size dependence}
As in the case of thin film systems, where the exchange bias field $H_{eb}$ is found to depend both on the thickness of the FM and AFM layers, EB effects in nanoparticles should depend on the particle size (core diameter $D_{C}$) and the thickness of the AFM shell $D_{Sh}$. 
The dependence of $H_{eb}$ on the particle core size should be similar to that on the thickness of the FM layer in thin film systems and, therefore, $H_{eb}$ should increase when reducing the particle size $H_{eb}\sim 1/D_C$. This trend was first reported in oxide passivated Co particles \cite{Gangopadhyay_nanostr92,Gangopadhyay_jap93} in the size range of 5-35 nm and later confirmed by Peng and coworkers \cite{Peng_pssa99,Peng_apl99,Peng_prb00,Sumiyama_stam05} on oxide coated Co/CoO particles with sizes 6-13 nm obtaining EB fields as large as 10.2 kOe for the smallest particles and a coercivity of 5 kOe. This has also been observed in oxygen passivated Fe particles with diameters of 6-15 nm \cite{DelBianco_prb02,DelBianco_jmmm03}, and in Fe/$\gamma$-Fe$_2$O$_3$ particles \cite{Zheng_prb04,Bomati_Small06}.
Moreover, a critical particle size below which EB is absent for any ratio of ferromagnetic and antiferromagnetic constituents has been reported \cite{Dobrynin_apl05} for Co/CoO 3 nm nanoparticles embedded in Al$_2$O$_3$. The reason is that, due to the large surface-to-volume ratio below the critical size, the exchange energy at the FM-AFM interface becomes smaller than both the effective Zeeman energy of the FM and the anisotropy energy of the AFM. In a later study on a sample consisting of 2.5 nm Co clusters embedded in a CoO matrix, the same authors performed a more complete study for samples with different oxide layer thicknesses \cite{Dobrynin_prb06}.
In fact, some authors have also reported an upper critical size (40 nm for the CoNi/CoO partices embedded in PVC of Ref. \onlinecite{Vazquez_phab04}) for the observation of EB. In another study of CoFe$_2$O$_4$ particles \cite{Mumtaz_jmmm07} with diameters 15-48 nm a nonmonotonic size dependence of $H_{eb}$, similar to what is observed in $H_C$, has been observed (although at much higher temperature of 77 K), with an increase with particle size up to a peak at around 27 nm followed by a subsequent decrease and vanishing for 40 nm particles.  
More recently, a study by Boubeta {\slshape et al.} \cite{Boubeta_prb06} on oxidized Fe particles with diameters ranging from 5 to 13 nm have confirmed the disappearence of EB  below a critical diameter of 5 nm and attributed this effect to the decreasing thickness of the spin-glass-like layer when decreasing the nanoparticle size. However, in the oxidized Fe particles studied by Ceylan {\slshape et al.} \cite{Ceylan_jap06}, the small particles ($7.5$ nm in diameter) were found to have much higher $H_{EB}$ than the big ones ($13$ nm in diameter), probably due to the increased relative effect of the AFM shell and the more amorphous structure of the shell in the smallest particles. 

\subsection {Shell thickness dependence}
Fewer studies have focused on the role played by the shell thickness, since the formation of oxidized phases cannot be easily controlled independently of the core size. As indicated by some models of EB for thin films \cite{VanderZaag_jap96,Ambrose_jap98,Ambrose_jap98b,Xi_prb00,Xi_prb00b,Ali_prb03,Ali_prb03b,Radu_thesis05,Radu_jpcm06}, there should be a minimum critical shell thickness for the observation of a loop shift, since the anisotropy energy per unit area of the AFM has to be larger than the interfacial exchange energy for EB to exist. Above this limiting thickness, $H_{eb}$ should increase with $D_{Sh}$ up to a critical shell thickness above which it would become independent of $D_{Sh}$. This has been partially corroborated by several works on nanoparticles of different compositions in which samples prepared by the same technique but different degrees of oxidation were compared \cite{Hou_jmmm98,Peng_jap02,Jeyadevan_ie02,Baker_jmmm04,Baker_jap04,Morel_jap04,Riveiro_apl05,Tracy_prb05,Dobrynin_prb06}. Moreover, the critical shell thickness in nanoparticles should depend on the anisotropy of the AFM as was first established in bilayers by Lund et al. \cite{Lund_prb02}. 

\subsection {Training effects}
A less studied effect, first described for thin films \cite{Paccard_pssb66} but also observed in nanoparticles, is the so-called training effect, which is observed when the hysteresis loop are succesively repeated a number of times $n$ after FC. $H_{eb}$ gradually decreases with $n$ in thin films, reflecting the deviation of the AFM spin structure at the interface layer from its equilibrium configuration \cite{Binek_prb04,Binek_prb05}. The relaxation of the frozen spins along the cooling field direction reduces the effective pinning energy, resulting in a decrease of $H_{eb}$ with the number of field cycles. Moreover, the bias field increase with increasing sweep rate of the magnetic field has been described by a dynamically generalized theory based on triggered relaxation, in excellent agreement with the
experiments \cite{Binek_prl06,Sahoo_jap07,Malinowski_apl07}.
A quantitative explanation based on the Kolmogorov-Avrami model describing the dynamics of AFM layers \cite{Xi_prb07} seems to describe correctly experimental data on the $H_{eb}(t)$ dependence. Also the symmetry of the anisotropy in the AFM seems to be crucial for the understanding of training effects \cite{Hoffmann_prl04}.

In core/shell nanoparticles, this training effect is characterized by a decrease of the coercive field on the descending field branch of the loop, whereas the ascending branch is usually retraced on succesive cyclings. Moreover, the training rate seems to depend strongly on the properties, namely the AFM or ferrimagnetic character, of the oxide shells. Thus, whereas in some Co/CoO particle systems the training is more pronounced after the second cycle \cite{Peng_prb00,Dobrynin_prb06,Tracy_prb06}, in some Fe/Fe oxide particle systems\cite{Peng_jap02}, the training effect is only decreased to about 89\% after the 14$^\mathrm{th}$ cycle (see also Refs. \onlinecite{Zheng_prb04,Trohidou_jmmmun}). Clearly related to training effects is  the observation of aging effects on $H_{eb}$ when the hysteresis loops are measured at different waiting times after the cooling field is applied \cite{DelBianco_jmmmun}.

\subsection {Temperature dependence}
Of course, both $H_{eb}$ and $H_{c}$ are thermal dependent quantities. Since the AFM or ferrimagnetic magnetic order at the particle shell, which is at the origin of the existence of EB, is degraded by temperature, EB should disappear when approaching the ordering temperature of the shell $T_N$, which is lower than the Curie temperature $T_C$ of the FM core. In fact, for most experimental systems, EB disappears at a so-called blocking temperature $T_B$ lower than $T_N$, although this is not necessarily true for $H_c$, for which finite values higher than those obtained after ZFC are usually observed up to $T_N$ \cite{Radu_prb03}. 

For thin films, it has been argued that the difference between $T_B$ and $T_N$ depends on the AFM layer thickness and is not related to finite-size effects on $T_N$ \cite{VanderZaag_prl00}.
However, in particle systems, this has been attributed to the SP behavior of the AFM oxide shell at a temperature lower than the $T_N$ of the shell, which might be composed of very small crystallites \cite{Gangopadhyay_jap93,Wen_jmmm04,Spasova_jm04}.  
%
With respect to the exact $T$ dependence, in thin films, linear dependencies of both quantities are usually observed \cite{Gruyters_prb01,Miltenyi_prl00,Radu_prb03} in accordance with the random field model of Malozemoff \cite{Malozemoff_prb87,Malozemoff_prb88,Malozemoff_jap88}. This is not always the case for core/shell nanoparticles, for which faster than linear decays of $H_{eb}$ and $H_{c}$ have been reported for Co/CoO particles \cite{Gangopadhyay_jap93,Peng_prb00,Wen_jmmm04}, although quasi-linear dependencies are also found \cite{Normile_jap06}. 
A law of the kind  $H_{eb}(T)=H_{eb}(0)(1-T/T_N)^n$ with $n=3/2$ has been shown \cite{Wiedwald_thesis,Spasova_jm04,Wiedwald_phtr05} to fit experimental data on Co/CoO particles, which is in accordance with the predictions of a model for polycrystalline bilayers \cite{Stiles_prb99b} that takes into account the thermal instability of the AFM shell. It must also be remembered that, when dealing with nanoparticle systems, other factors apart form the structural ones, intrinsic to the particle, such as the volume distribution, randomness of the anisotropy axes and the existence interparticle interactions \cite{Dormann_acp97} may influence the thermal dependence of both $H_{eb}$ and $H_{c}$.

\subsection {Cooling field dependence}
There is no general trend for the dependence of $H_{eb}$ on the cooling field magnitude in layered systems. Depending on the details of the microscopic structure of the interface and the AFM layer and the preparation conditions, both a slight decrease  \cite{Moran_jap96} or increase \cite{Ambrose_jap98} of $H_{eb}$ with increasing $T$ have been reported. 
However, some systems \cite{Nogues_jmmm99} display loops shifts towards positive field values instead of to negative fields for large cooling fields. This effect has been argued to be possible when the coupling at the interface is AFM. Estimations of the crossover field have been given \cite{Nogues_prl96,Hong_prb98,Kiwi_epl99,Kiwi_ssc00}, and experiments have also proved the validity of the hypothesis in several bilayered systems \cite{Nogues_prb00,Radu_prb03,Yang_prb05,Henry_prb06,Arenholz_apl06,Huang_apl07}.

Field cooling dependencies have been reported only recently in core/shell nanoparticles. In CoFe$_2$O$_4$ particles \cite{Mumtaz_jmmm07}, $H_{eb}$ has been found to increase with the cooling field for values of $H_{FC}$ up to 5000 Oe while, for higher fields, a slight decrease is observed accompanied by a decrease in the vertical loop shift.
On the other hand, while for Co/CoO nanoparticles \cite{Zhou_apa05} $H_{eb}$ continue to increase for fields up to 5 T with values of the order of 1-2 kOe at 300 K, for Fe/FeO nanoparticles \cite{DelBianco_prb04,DelBianco_jmmm05,Fiorani_jmmm06}, $H_{eb}$ presents a maximum at a field cooling value around 5 kOe which increases with decreasing $T$. In this case, $H_{eb}$ decreases with further increasing the cooling field, reaching a value of only 250 Oe at 5 T and 5 K. The authors argued that the appearance of the maximum is due to the glassy magnetic nature of the oxide phase at the shell, which might be destroyed by increasing magnetic fields or temperatures. A similar behaviour has also been reported in phase-separated LSCO perovskite \cite{Tang_jap06}. A clear-cut interpretation for these systems is still lacking.

\subsection {Asymmetry of the hysteresis loop}
Another commonly observed feature in bilayers is an asymmetry between the descending and increasing field branches of the loops after FC, which has been related to different magnetization reversal mechanisms in each of the branches. While in the descending field branch reversal takes place usually by uniform rotation, in the increasing field branch, reversal by nucleation and propagation of domain walls or non-uniform structures seems to be the dominant mechanism. 
Different techniques, sensitive to microscopic magnetic configurations of the FM and AFM, have confirmed these different reversal mechanisms. First studies on this issue were performed in FeNi/FeMn films by magneto-optical methods \cite{Nikitenko_prl00,Gornakov_prb06} and in MnF$_2$/Fe \cite{Fitzsimmons_prl00,Leighton_prl01} and CoO/Co bilayers \cite{Gierlings_prb02,Radu_apa02,Radu_prb03,Paul_prb06} by polarized neutron reflectometry.
Later on, also X-ray photoemission microscopy has been used in Fe/MnPd films \cite{Blomqvist_prl05}, time-resolved Kerr magnetometry in FeF$_2$/Fe bilayers \cite{Engebretson_prb05} and neutron scattering in patterned Co/CoO nanostructures \cite{Temst_jmmm06}. Recently, the origin of asymmetric loops in some particular systems has been ascribed to the competition between the FM and the interfacial FM-AFM exchange anisotropies \cite{Chung_prb05,Camarero_prl05} and different reversal processes in both loop branches have also been revealed by SXRMS and techniques in perpendicularly coupled exchange coupled films \cite{Camarero_apl06} and by MOKE in Fe/MnF$_2$ bilayers \cite{Tillmanns_apl06}.

Asymmetries in the shape of the hysteresis loops of core/shell particles are also evident in some systems, but, in this case, present experimental techniques cannot easily give information about the microscopic mechanisms involved in the reversal processes because of the particle size dispersion always present in samples. For this purpose, experiments being able to measure magnetic properties of a single nanoparticle (in the spirit of those performed by Wernsdorfer and coworkers \cite{Bonet_prl99,Jamet_prl01,Thirion_NMat03}) would help to clarify this controversial issue.

\subsection {Vertical loop shifts}
In some systems, shifts along the magnetization axes have also been reported \cite{Nogues_prb00,Liu_apl04} that have been related to induced magnetic moments. This vertical shift depends on the cooling field (it may be negative for low $H_{FC}$ and positive for large $H_{FC}$) and the microstructure of AFM layer. Recently, X-ray magnetic circular dichroism (XMCD) experiments on Ni/FeF$_2$ bilayers have proved that the vertical shift is due to the existence of uncompensated Fe pinned moments in the AFM \cite{Arenholz_apl06}.
Huang and co-workers \cite{Huang_apl07} have observed linear dependence of the exchange field on the magnetization shift in ZnCo$_0.07$O/NiO layers, proving the role of uncompensated pinned spins on the observation of the effect. 
Some core/shell nanoparticles also display this phenomenology. 
Vertical shifts have been reported\cite{Gangopadhyay_jap94} for Ni/NiO \cite{Yao_nanostr95,Yao_msea96}, Co/CoO \cite{Zhou_apa05,Dobrynin_apl05,Dobrynin_prb06,Tracy_prb06}, Fe/Fe$_2$O$_3$ \cite{Zheng_jap04,Zheng_prb04,Ceylan_jap06} and in milled Fe/MnO$_2$ \cite{Passamani_jmmm06} particles, with values much higher than those reported for bilayers. 
The linear dependence of the vertical shifts measured at different temperatures on H$_{eb}$ found in \onlinecite{Zheng_jap04,Zhou_apa05} indicates that the vertical shifts are proportional to the number of net frozen spins. A nonmonotonic dependence of the shifts on the particle size and cooling field, in agreement with that found for $H_{eb}$, has been reported by Mumtaz et al. \cite{Mumtaz_jmmm07}.

\subsection {Nature of the interface coupling}
Recently, several spectroscopic techniques have provided insight on the structure and magnetic behavior of the interface spins at a microscopic level, demonstrating the crucial role played by uncompensated interfacial spins on EB in several bilayered thin film systems \cite{Nolting_nature00,Ohldag_prl01,Ohldag_prl03,Kappenberger_prl03,Blomqvist_prl05,Valev_prl06,Camarero_apl06} and also demonstrating unambiguously the existence of domain walls in the FM parallel to the AF/FM interface \cite{Morales_apl06}.
Similar techniques applicable to nanoparticles such as X-ray absorption and XMCD have also been used to study Fe oxide passivated iron nanoparticles \cite{Fauth_jap04}. The relative sign of the metal and oxide related dichroism allows to conclude that the coupling across the interface is FM. This finding is opposed to the situation at the Fe(110)/Fe$_3$O$_4$ interface, where an AFM coupling was found \cite{Kim_prb00}. 
Presence of uncompensated Co magnetic moments at the interface of a $2$–$2.5$ nm CoO shell surrounding a metallic fcc-like 7-8 nm Co core was also evidenced by XMCD \cite{Wiedwald_prb03,Wiedwald_phtr05}.

\subsection {Other recent observations}
In this last subsection, we would like to mention some very recent experimental observations in core/shell nanoparticle systems which have given evidences of new phenomenology not mentioned in the previous subsections and that we think will estimulate further studies both from the experimental and theoretical point of views. 
Tracy {\slshape et al.} \cite{Tracy_prb06} have reported an investigation of the role of defects on the magnetic properties of Co/CoO nanoparticles in which, by measuring magnetization and thermoremanence curves under ingenious FC protocols with intermediate field reversals, they are able to show that the defect moments freeze at low temperature and have a distribution of melting temperatures and that they dominate EB at low temperature, exhibiting also a thermal memory effect. The role of dilution on the AFM have also been studied in bilayers \cite{Papusoi_jap06,Fecioru_condmat07}. Both experimental and simulation results confirm an enhancement of $H_{eb}$ with increasing defect concentration.
Nogués {\slshape et al.} \cite{Nogues_prl06} have demonstrated that the magnetic
properties of Co/CoO nanoparticles embedded in an Al$_2$O$_3$ matrix, depend strongly on the in-plane coverage, even in the diluted regime. In particular, the authors have found that both $H_C$ and $H_{eb}$ radically increase with increasing coverage. The experiments allow the authors to conclude that these observations cannot be accounted by dipolar interactions between the cores and should be attributed to shell mediated interactions when particle become in contact. This would also help to explain the scatter of values for $H_C$ and $H_{eb}$ found in the literature.
A study of CoO granular films deposited on layered FM structures by Gruyters \cite{Gruyters_prl05} have shown that EB in this system can be explained by the spin-glass-like state in the nanoparticles constituting the CoO film without the need for core/shell structure. These results show that pinning effects in EB systems are not only related to uncompensated spins, but may arise due to a frozen state in the AFM similar to a spin-glass. Moreover, the deduced unusually large uncompensated magnetization has no simple quantitative relation to $H_{eb}$, a fact that requires further theoretical developement in order to be understood. The same author has proposed a model \cite{Gruyters_epl07}, based on the random magnetic anisotropy of CoO nanoparticles, according to which the observation of EB can be attributed to an interaction between the AFM order and uncompensated spins in the AFM material without explicitly invoking the exchange coupling to a FM.
Another study that will hopefully provide a new direction for studies of EB is that by Ali and co-workers \cite{Ali_nature07} on a Co/CuMn bilayer system, which has evidenced the possibility of observing most of the EB associated phenomenology using a spin-glass material instead of conventional AFM. A striking difference form FM/AFM bilayers a change in sign of the bias field just below the blocking temperature has been found in this system, indicating that the indirect RKKY exchange within the pinning layer may account for the observed effects. One may wonder if core/shell particles with similar morphologies could also give surprising new effects. 

\section{Models and simulations}
Some microscopic models for bilayers have undertaken calculations of EB fields under certain assumptions \cite{Takano_prl97,Scholten_prb05}, numerical studies based on a mean field approach \cite{Almeida_prb02} or Monte Carlo (MC) simulations \cite{Nowak_prb02,Suess_prb03,Lederman_prb04} making different assumptions about the interface. However, only very recently, some works partially addressing the EB phenomenology in nanostructures have been published \cite{MejiaLopez_prb05,Eftaxias_prb05}.  

\subsection{Model of core/shell particle}
\begin{figure}[t]
\includegraphics[width=0.7 \columnwidth, angle= -90]{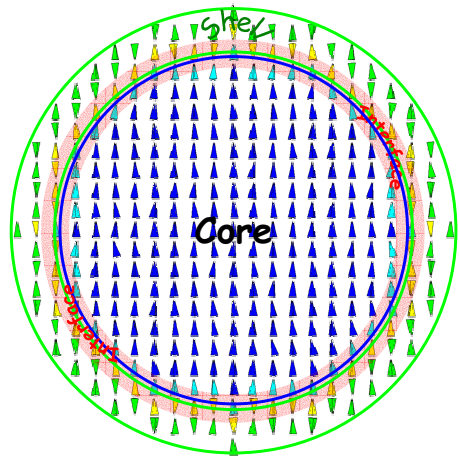}
\caption{\label{Fig_1}(Color online) Schematic drawing of model of a core/shell nanoparticle of total radius $R$ used in the MC simulations. The spins sit on the nodes of a sc lattice. The AFM shell has width $R_{Sh}$ (green and yellow spins) and the FM core (blue spins) a radius $R_C= R-R_{Sh}$. The core/shell interface (light blue and yellow spins) is formed by the core (shell) spins having nearest neighbours on the shell (core).}
\end{figure}

In order to understand what is the microscopic origin of all the phenomenology associated to EB effects presented in the preceding section, we have developed a model for a single nanoparticle with core/shell structure which captures the main ingredients that are believed to be necessary for the observation of EB. A schematic drawing of the particle is shown in Fig. \ref{Fig_1}. Atomic spins are considered to sit on the nodes of a sc lattice and the particle is buildt by considering the spins inside a sphere of radius $R$ (measured in multiples of the unit cell dimensions $a$) centered in on of the lattice nodes. Three regions are distinguished inside the particle: a core with radius $R_C$, a shell of thickness $R_{Sh}=R-R_C$ and the core/shell interface that is formed by the core (shell) spins having nearest neighbours on the shell (core). 
In most of the results presented in the following, we have considered a fixed particle size $R= 12 a$ an a shell of thickness $R_{\mathrm{Sh}}= 3 a$. Taking $a= 0.3$ nm,  such a particle corresponds to typical real dimensions $R\simeq 4$ nm and $R_{\mathrm{Sh}}\simeq 1$ nm and contains $5575$ spins, of which $45$ \% are on the surface. 
Since we are interested in studying magnetic properties observed in real core/shell particles, we will consider that the core of the particle is made of a FM material and that the outer shell is an AFM. Different characteristic microscopic parameters, such as exchange and anisotropy, will be considered in the three regions, with fixed values at the core and shell regions and that will be varied at the interface in order to study what is its specific role in establishing EB properties. 

To account for the finite values of anisotropy in real systems, we have considered a model of Heisenberg classical spins $\vec{S}_i$, interacting according to the following microscopic Hamiltonian 
\begin{eqnarray}
\label{Eq1}
{ H}/k_{B}= 
-J_{\mathrm C}\sum_{\langle  i,j\rangle\in\mathrm{C}}{\vec S}_i \cdot {\vec S}_j   
-J_{\mathrm S}\sum_{\langle  i,j\rangle\in\mathrm{Sh}}{\vec S}_i \cdot {\vec S}_j \nonumber\\  
-J_{\mathrm{Int}}\sum_{\langle  i\in\mathrm{C},j\in\mathrm{Sh}\rangle}{\vec S}_i \cdot {\vec S}_j 
-k_C\sum_{i\in \mathrm{C}}(S_i^z)^2\nonumber\\
-k_S\sum_{i\in \mathrm{Sh}}(S_i^z)^2  	
-\sum_{i= 1}^{N} \vec h\cdot{\vec S_i} \ .
\end{eqnarray}
The first three terms describe the nearest-neigbor exchage interactions between the spins with different values of the exchange constants at the different particle regions. Core spin are FM with $J_{\mathrm{C}}>0$, whereas spins in the shell are AFM with $J_{\mathrm{S}}<0$. The values of these constants will be kept constant and fixed arbitrarily to $J_{\mathrm{C}}=10$ and $J_{\mathrm{Sh}}=-0.5 J_{\mathrm{C}}$, which just fix the Curie temperature of the FM to $T_C= 29$ K and the Neél temperature of AFM to $T_N= 14.5$ K, a value lower than $T_C$ as is the case in most oxides with respect to their native materials.
Finally, for spins the exchange constant at the interaface $J_{\mathrm{Int}}\lessgtr 0$ will be allowed to vary between $0$ and $\pm J_C$ in order to study the role played by the coupling across the core/shell interface on magnetic properties.
\begin{figure}[t]
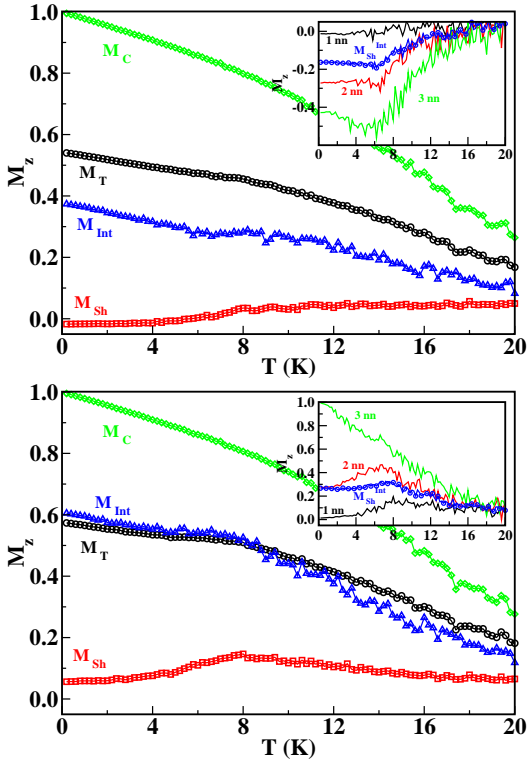
 
\centering 
\includegraphics[width=0.8\columnwidth ]{FC_Js-05_Ji-05_nn.eps}
\includegraphics[width=0.8\columnwidth ]{FC_Js-05_Ji+05_nn.eps}
\caption{(Color online) Thermal dependence of the normalized magnetizations of a core/shell particle when cooling form a disordered state at $T>T_N$ down to $T = 0.1$ in the presence of an external magnetic field $h_{\mathrm FC}= 4$ K. The values of the exchange coupling at the interface are (a) $J_{\mathrm{Int}}= -0.5 J_{\mathrm{C}}$ and (b) $J_{\mathrm{Int}}= +0.5 J_{\mathrm{C}}$.
The different curves correspond to the contributions of the core $M_{\mathrm{C}}$, shell $M_{\mathrm{Sh}}$ and interface $M_{\mathrm{Int}}$ spins to the total magnetization $M_{\mathrm{T}}$. 
Insets display the contributions of only the interfacial shell spins to $M_{\mathrm{Int}}$ ($M_{\mathrm{Sh}}^{\mathrm{Int}}$, in circles) and, among these, the ones having 1 (black), 2 (red) and 3 (green) nearest-neigbors in the core. 
}
\label{FC_fig}
\end{figure}

The fourth and fifth terms correspond to the on-site uniaxial anisotropy with $k_C$ and $k_\mathrm{Sh}$ the values of the anisotropy constants at the core and at the shell. They can be related to values in real units through the correspondence 
\begin{eqnarray}
k_\mathrm{C}=\frac{K_\mathrm{C}V}{N_\mathrm{C}},\ k_\mathrm{Sh}=\frac{K_\mathrm{Sh}S}{N_\mathrm{Sh}}
\end{eqnarray}
, where $K_\mathrm{C}$ and $K_\mathrm{Sh}$ are the anisotropy constants in units of energy per unit volume ($V$) or surface ($S$) of the particle.  
The value of $k_\mathrm{C}$ will be fixed to $k_\mathrm{C}=1$ K, which just sets the value of the anisotropy field of the FM core, whereas the anisotropy at the AFM shell has to be higher than that in the core as required to pin the AFM spins during the hysteresis loops so that EB is observed. Therefore, we fix $k_\mathrm{Sh}= 10$ K, which is which also in agreement with the reported enhanced surface anisotropies due reduced local coordination at the outer particle shells \cite{Tronc_jmmm03,Morales_chemmat99}.
Finally, the last term is the Zeeman energy coupling to an external magnetic field $H$, where $h=\mu H/k_B$  denotes the field strength in temperature units, with $\mu$ the magnetic moment of the spin.
\begin{figure}[tbp] 
\centering 
\includegraphics[width=0.99\columnwidth,angle= -90]{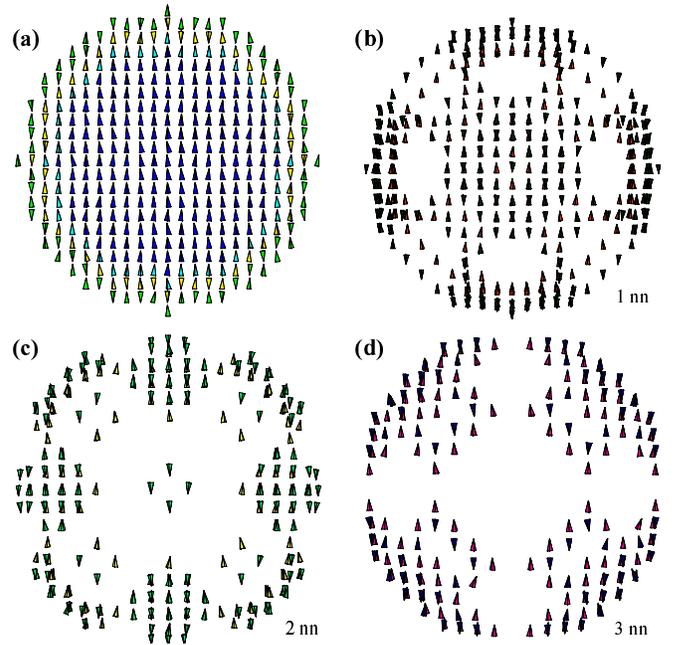}
\caption{(Color online) (a) Spin configuration of an equatorial cut of the particle attained after the field cooling process described in Fig. 1. Core spins are dark blue, spins at the shell are green while inteface core and shell spins have been colored in light blue and yellow. (b-d) Configurations of the core (shell) spins at the interface having 1, 2 or 3 nearest neighbors in the shell (core). [Reprinted with permission from Ref. \onlinecite{Iglesias_phyab06}, O. Iglesias et al. Physica B {\bfseries 372}, 247 (2006). Copyright @Elsevier B. V.]
}
\label{FC_Confs_fig}
\end{figure}

Based on this Hamiltonian, we have performed Monte Carlo simulations using Metropolis algorithm. As for the spin updates, we use a combination of the trial steps which has proved useful for Heisenberg spins with finite anisotropies as described elsewhere \cite{Hinzke_cpc99,IglesiasphaB04}. 

\subsection{Field cooled states}
\label{FCS_Sec}
First, we will study the magnetic state of the particle after a field cooling process with the purpose to characterize the magnetic order induced on the interfacial spins. Our protocol to simulate the field cooling process is as follows. We start the simulations from a high temperature $T_0> T_N$ disordered state in which the spins are pointing in random directions with zero net magnetization. The temperature is then reduced in constant steps $\delta= 0.1 $K down to the final temperature $T= 0.1$ K in the presence of a magnetic field $h_{\rm{FC}}= 4$ K applied along the easy-axis direction. At each temperature, the magnetization is averaged over a number of 10000 MC steps after 10000 MC steps used for thermalization, using the usual heat bath dynamics for continuous spins. 

As an example, the thermal dependence of the normalized magnetization along the field direction is shown in Fig. \ref{FC_fig} for a particle with AFM or FM interface coupling $J_{\mathrm{Int}}=\mp 0.5 J_{\mathrm{C}}$. In this figure, the contributions of the spins in the core ($M_C$), in the shell ($M_{\mathrm{Sh}}$) and at the interface ($M_{\mathrm{Int}}$) to the total magnetization $M_T$ have been plotted separately. 
As it can be seen in the main panels, during the cooling process, the core spins progressively order ferromagnetically as indicated by the increase of $M_C$ towards 1. At the same time, as $T$ is reduced below the Neél temperature of the shell, the AFM order is also established in the shell spins, although a finite value of $M_{\mathrm{Sh}}$ remains at the lowest temperature due to the noncompensation between sublattices caused by the finite-size and spherical shape of the particle. Most importantly, independently of the nature of the coupling between the core and shell spins, the interfacial spins are not compensated, as indicated by the finite magnetization attained at low $T$, which, of course, is lower in the AFM case ($M_{\mathrm{Int}}= 0.37)$ than in the FM one ($M_{\mathrm{Int}}= 0.605$. 
\begin{figure}[tbp]
\includegraphics[width=\columnwidth]{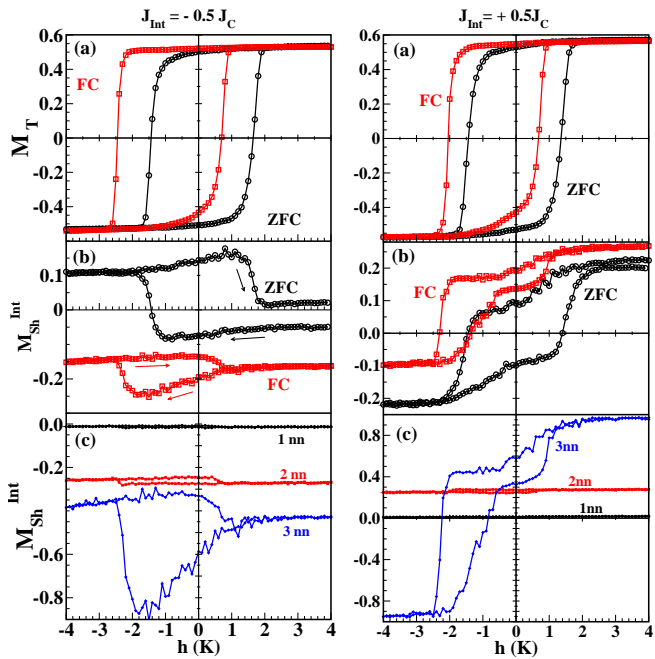}
\caption{\label{Fig_FCH}(Color online) Hysteresis loops for a particle with radius $R=12\,a$  obtained from a ZFC state and after FC down to $T= 0.1$ in a field $h_\mathrm{FC}$ for $J_\mathrm{Sh}= -0.5 J_\mathrm{C}$ and $J_\mathrm{Int}=-(+) 0.5 J_\mathrm{C}$ in the left (right) column. Panels (a) display the total normalized magnetization component along the field direction. Panels (b) show the normalized contributions of the shell spins at core/shell interface to the total magnetization of the loop. Panels (c) show the contribution of the interfacial spins at the shell to $M_\mathrm{Sh}^\mathrm{Int}$ having 1, 2 or 3 nn in the core.}
\end{figure}

In order to gain deeper understanding on the origin of this net interface magnetization, first notice that the interfacial spins at the core are all pointing in the field direction after the FC process. Therefore, uncompensated moments must be originated at the shell interface. We show in the insets of Fig. \ref{FC_fig} the contributions to $M_{\mathrm{Int}}$ of the interfacial spins at the particle shell in the curve labeled $M_{\mathrm{Sh}}^{\mathrm{Int}}$ (in circles). Comparing the insets in panels (a) and (b), we see that the sign of the net magnetization at the shell interface is in accordance with sign of the interface coupling. Further inspection of the contributions of spins having different number of nn in the core, allows us to conclude that the aligning effect of the cooling field is more effective for the spins with lower number of nn in the AFM case and for those with more nn in the FM case. It can also be seen that the major contribution to the net interface magnetization comes from the shell spins with 3 nn in the core, independently of the sign of $J_{\mathrm{Int}}$.
As it can be concluded from the preceding observations, the geometric structure and magnetic ordering of the interface in a core/shell nanoparticle is more intricate than in the case of FM/AF coupled bilayers due to the roughness inherent to the geometry of the interface. At difference from bilayers, interfacial spins may have different number of neighbors depending on their position and, therefore, the interface spins at the shell present regions with either local compensated or uncompensated magnetic order.

\subsection{FC hysteresis loops}

In order to study the phenomenology associated to EB effects, we have also performed simulations of hysteresis loops following a protocol that mimics the experimental one: configurations obtained at the lowest temperature after the FC process described in the preceding section are used as the starting state and then the hysteresis loops are recorded by cycling the magnetic field from $h= 4$ K to $h= -4$ K in steps $\delta h= -0.1$ K and the different quantites averaged during $200$ MC steps per spin at every field after other initial $200$ MC stepd are discarded for thermalization. Hysteresis loops obtained from a zero field cooled (ZFC) state have also been simulated starting from a demagnetized state at the measuring temperature, then following the first magnetization curve up to $h= 4$ K and, finally, performing the hysteresis loop as described before.

Typical ZFC and FC hysteresis loops are shown in Fig. \ref{Fig_FCH} (upper panels) for two values of the interface coupling $J_\mathrm{Int}/J_\mathrm{C}= -0.5, +0.5$. Compared to the loops obtained from ZFC state, the loops obtained after FC are shifted towards negative field values and have slightly increased coercivity (see Figs. \ref{Fig_1}a), independenly of the sign of the interfacial exchange coupling. The values of the coercive fields for the decreasing and increasing field branches will be denoted by $h_{\mathrm C}^-$ and $h_{\mathrm C}^+$, respectively. Therefore, the coercive field and the EB fields are defined as $h_\mathrm{C}= (h_\mathrm{C}^+ - h_\mathrm{C}^-)/2$ and $h_\mathrm{eb}=(h_\mathrm{C}^+ + h_\mathrm{C}^-)/2$, respectively.
The origin of the shift in the FC case can be better understood by looking at the contribution of interface spins belonging to the shell, $M_{\mathrm Sh}^{\mathrm Int}$, to the total magnetization as displayed in the middle panels of Fig. \ref{Fig_FCH}. 

As we have previously revealed by the detailed inspection of the microscopic configurations attained after FC, the interfacial spins at the shell acquire a negative ($J_{\mathrm{Int}}<0$) or positive ($J_{\mathrm{Int}}>0$) net magnetization after FC, in both cases higher than the one attained after ZFC, although more pronounced for the negative coupling case.
This net magnetic moment, induced by the geometrical symmetry breaking and the alignment of groups of spins into the field direction, generates local fields on the core spins that point into the same direction as the external field, causing the shift of the hysteresis loops. 

To further support this observation, we note that the hysteresis loops are shifted by the same amount but towards the positive field axis when cooling in a field applied in a direction negative with respect to the measuring field (see for example the dashed lines in Fig \ref{Fig_Asy} for $J_{\mathrm{Int}}/J_{\mathrm{C}}= -0.5, -1$).
These observations reflect that, after the FC process, a fraction of the interfacial spins ($\approx$15 \% of the interface spins at the shell) have been pinned along a direction compatible with the core/shell exchange interaction, as corroborated also by the vertical shifts in the $M_\mathrm{Sh}^\mathrm{Int}$ loops (to be commented below). This is no longer true for the ZFC case, for which a high fraction of interfacial spins follows the reversal of the FM core, as reflected by the change in sign of $M_\mathrm{Sh}^\mathrm{Int}$ along the hysteresis loop.
Moreover, FC hysteresis loops obtained for the same particles but without increased anisotropy at the AFM shell (performed setting $k_\mathrm{Sh}= 1$) display no EB but, instead, have increased coercive fields compared to ZFC loops. In this case, no interfacial shell spins are pinned and, during reversal, they are dragged by the core spins due to the dominance of exchange coupling over anisotropy energy. This observation demonstrates that high anisotropy AFM are required to obtain exchanged biased loops.

It turns out that disorder and frustration at the surface induced by radial anisotropy and finite-size effects alone are not enough to produce sizable loops shifts as simulations performed for particles with no AF shell demonstrate \cite{Iglesiasprb01}. 

\subsection{Quantifying $h_{eb}$: microscopic origin of EB}

One of the most controversial points in EB concerns the evaluation of the loop shifts from a model of the system at hand. Different theories and models usually predict EB shifts that differ by orders of magnitude from that measured experimentally. An archetypical example is the expression first derived by Meiklejohn and Bean (MB) \cite{Meiklejohn_pr56,Meiklejohn_pr57,Meiklejohn_jap62} for a bilayer that reads
\begin{eqnarray}
	H_{eb}= \frac{J_{eb}}{\mu_0 M_F t_F} \ ,
\end{eqnarray}
where $J_{eb}$ is the interfacial exchange energy per unit area and $M_F$, $t_F$ are the magnetization and thickness of FM layer, respectively. Although this expression describes correctly the linear decrease of $H_{eb}$ with $t_F$, it fails in the quantitative prediction of most of the measured loop shifts, the reasons being, essentially, that the FM/AFM interface is supposed to be fully uncompensated and ideally smooth and that the  AFM is considered to be single domain with spins that remain unchanged during the reversal of the FM. Other models based on refined versions of the MB model gave improved expressions for $H_{eb}$ that agreed more reasonably with experimental values in some layered systems. 
Let us briefly recall that the model by Malozemoff \cite{Malozemoff_prb87,Malozemoff_prb88,Malozemoff_jap88}, that incorporated the roughness of the interface as a random field acting on the FM layer and a model by Mauri \cite{Mauri_jap87} that, following the pioneering work by Neél \cite{Neel_Ann67}, accounted for the possibility of domain wall formation in the AFM, gave modified expressions for the EB field of the kind $H_{eb} \sim \frac{\Delta \sigma_{AF}}{\mu_0 M_F t_F}$ ($\Delta \sigma_{AF}$ being the domain wall energy density in the AFM), which result in reduced values with respect to the MB model (see also the models by Kiwi at al. \cite{Kiwi_epl99,Kiwi_ssc00,Kiwi_jmmm01} and Stamps and co-workers \cite{Stamps_jpd00,Kim_prb05}). 

In spite of the profusion of models presented above, none of them takes into account the evolution of the spin structure of the FM and the AFM along the hysteresis loops and this is the reason for their lack of agreement with experiments. 
More microscopic approaches such as the work by Takano et al. \cite{Takano_prl97}, in which, by calculating the density of interfacial uncompensated spins in permalloy/CoO bilayers, the authors predicted the correct magnitude of the exchange field, as well as
the observed inverse dependence on interfacial grain size, have been more successful.
More recently, a semi-quantitative account of the EB field magnitude has been presented in a simplified model for Co nanoparticles embedded in a CoO matrix \cite{Givord_jmmm05}.
In order to link the measured loop shifts to the microscopic details of the samples, 
Monte Carlo and micromagnetic simulations based on microscopic models \cite{Koon_prl97}  have proved useful. Among them, let us mention here that, to our knowledge, only the domain state (DS) model proposed by Nowak and collaborators \cite{Miltenyi_prl00,Nowak_prb02,Beckmann_prl03,Beckmann_prb06,Spray_jpd06} have been able to establish a numerical correspondence between $H_{eb}$ and microscopic parameters by proving that $H_{eb}$ is proportional to the irreversible domain state magnetization of the AFM interface layer $m_{IDS}$ as $H_{eb}=\frac{J_{Int}m_{IDS}}{l\mu_o\mu}$, where $l$ is the FM layer thickness and $\mu$ the atomic magnetic moment. 
\begin{figure}[t]
\includegraphics[width=0.9\columnwidth]{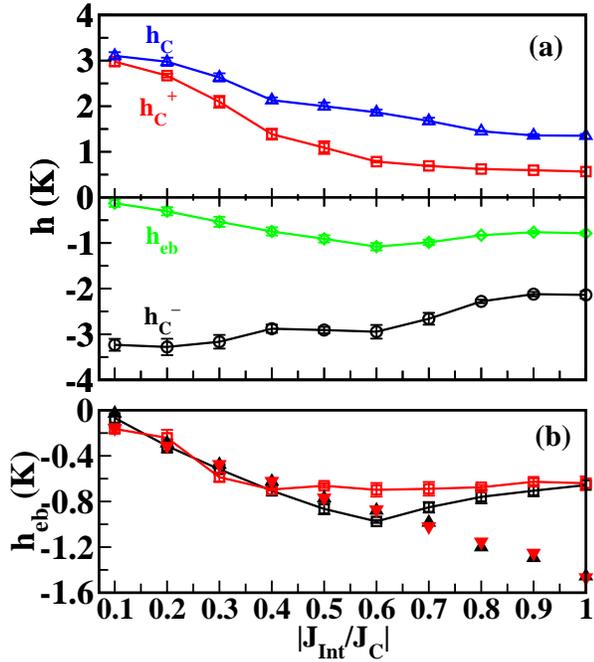}
\caption{\label{Fig_HebHc}(Color online) (a) Variation of the coercive fields $h_\mathrm{C}^-$, $h_\mathrm{C}^+$, $h_\mathrm{C}$ and the exchange bias field $h_\mathrm{eb}$ with the exchange coupling constant at the core/shell interface for $J_\mathrm{Int}< 0$ for a particle with $J_\mathrm{Sh}= -0.5 J_\mathrm{C}$.
(b) Variation of $h_{eb}$ with  $J_\mathrm{Int}< 0$ (open circles) and $J_\mathrm{Int}> 0$ (open squares). The exchange bias fields computed from Eq. \ref{Eq_eb} as described in the text are shown as filled symbols for $J_\mathrm{Int}< 0$ (down triangles) and $J_\mathrm{Int}> 0$ (up triangles).
}
\end{figure}

In the case of a core/shell nanoparticle, the analysis is more intricate due to the peculiarities of the core/shell interface as already commented in Sec. \ref{FCS_Sec}, and a more detailed analysis is needed. 
In order to elucidate the role played by the interface in establishing the EB effect, we have studied the variation of $h_\mathrm{C}^-$, $h_\mathrm{C}^+$, $h_\mathrm{C}$ and $h_\mathrm{eb}$ with the interface exchange coupling $J_\mathrm{Int}$, presented in Fig. \ref{Fig_HebHc}a for negative $J_\mathrm{Int}$ values.
With increasing $J_\mathrm{Int}$, both $h_\mathrm{C}^-$ and $h_\mathrm{C}^+$ decrease in absolute value, although they seem to reach a constant value when approaching $|J_\mathrm{Int}|= J_\mathrm{C}$. As a consequence, a decrease in $h_\mathrm{C}$ and an increase in $h_\mathrm{eb}$ is observed, with a nearly linear dependence, at least for values of $|J_\mathrm{Int}|$ smaller than the exchange coupling at the shell $J_\mathrm{Sh}= -0.5 J_\mathrm{C}$. Similar linear dependencies have been found in the DS model and some other models of bilayers \cite{Mewes_apl04}. MC simulations of a cylindrical nanodot \cite{MejiaLopez_prb05}, also demonstrated an increase in $h_\mathrm{eb}$ with the scaled effective unidirectional anisotropy.
For both $J_\mathrm{Int}\gtrless 0$, the values of $h_\mathrm{C}$ and $h_\mathrm{eb}$ are very similar, as can be seen in Fig. \ref{Fig_HebHc}b.
With the increase of $|J_\mathrm{Int}|$, core spins become more coupled to the unpinned shell spins, therefore facilitating the magnetazation reversal with the subsequent decrease in the coercivity, an observation also found in micromagnetic simulations of a model of coupled bilayers with grains in the AFM, which exhibit random uniaxial anisotropy and are weakly exchange coupled \cite{Stiles_prb01,Kirschner_ieee03,Suess_jap03}.
At the same time, increasing $|J_\mathrm{Int}|$ while keeping the values of $J_{\mathrm C}$, $J_\mathrm{Sh}$ and $h_\mathrm{FC}$ constant, results in higher local exchange fields created by the uncompensated spins at the interface, causing an increase of the loop shift.
Notice, however, that, increasing $|J_\mathrm{Int}|$ above $J_\mathrm{Sh}$ do not result in a further increase of $h_\mathrm{eb}$, which seems to converge to a common value for both cases. The reason for this departure from linearity stems will be commented in the next sections.  
Finally, let us also mention that the values of the coercive and exchange bias fields obtained from simulations are within the correct order of magnitude when expressed in real units. For example, for $J_\mathrm{Int}/J_\mathrm{C}\in [-0.3,-0.5]$, we obtain $H_\mathrm{C}\approx 1.3-1$ T and $H_\mathrm{eb}\approx 0.27-0.43$ T, which are in agreement with typical values found in studies of oxidized nanoparticles \cite{Skumryev_nature03,Gangopadhyay_jap93,Luna_nano04,Peng_prb00,Punnoose_prb01,DelBianco_prb04,Zheng_prb04,Tracy_prb05,Nogues_prl06}.  

The proportionality of $h_\mathrm{eb}$ to $J_\mathrm{Int}$ should be taken as a hint for the microscopic origin of the loop shifts. As we have mentioned in previous paragraphs, the observed vertical displacements of the loops corresponding to the interface shell spins point to the existence of a net magnetization at the core/shell interface due to uncompensated pinned spins at the shell interface \cite{Nowak_prb02}. If this is the case, the coercive fields after FC can be thought as the sum of the ZFC coercive field $h_\mathrm{C}^0$ and the local field acting on the core spins due to the net interface magnetization of the shell spins, so that they may be computed as \cite{Scholten_prb05,Iglesias_prb05}
\begin{eqnarray}
	h_\mathrm{C}^\pm= h_\mathrm{C}^0 + J_\mathrm{Int} M_\mathrm{Int}^{\pm} \ ,
\end{eqnarray}
where $M_\mathrm{Int}^{\pm}=\sum_{i\in \{\mathrm{Int},{\mathrm{Sh}}\}} z_i S_i^z$ is the net magnetization of the interfacial shell spins at the positive (negative) coercive fields $h_\mathrm{C}^\pm$, and $z_i$ is the number of nearest neighbors of spin $i$. Therefore, the coercive and exchange bias fields can be written as
\begin{eqnarray}
\label{Eq_eb}	
	h_\mathrm{C}= h_\mathrm{C}^0+ J_\mathrm{Int} (M_\mathrm{Int}^{+}- M_\mathrm{Int}^{-})/2\\ \nonumber 
	h_\mathrm{eb}= J_\mathrm{Int} (M_\mathrm{Int}^{+}+ M_\mathrm{Int}^{-})/2\ .
\end{eqnarray}

These expressions establish a connection between the coercive fields and loop shifts observed macroscopically and microscopic quantities that, although may not be directly measured in an experiment, can be computed independently from the simulation results. 

The values of $h_\mathrm{eb}$ obtained by inserting the $M_\mathrm{Int}^{\pm}$ values extracted from the Fig. \ref{Fig_FCH}b in Eq. \ref{Eq_eb} are represented as filled symbols in Fig. \ref{Fig_HebHc}b, where we can see that the agreement with the $h_\mathrm{eb}$ values obtained from the hysteresis loop shift is excellent within error bars. 
Recent experiments by Morel et al. \cite{Morel_prl06} on Co particles embedded in MnPt films have observed a clear correspondence between the measured $H_{eb}$ and $M_{AFM}$, the magnetization induced in the AFM MnPt established by suitable FC procedures, which reinforce the validity of our model.
Only for $|J_\mathrm{Int}| > J_\mathrm{Sh}$, an increase in $|J_\mathrm{Int}|$ does not result in a further increase of $h_\mathrm{eb}$, as reflected by a departure from linearity implied by Eq. \ref{Eq_eb}, which means that the interfacial net magnetization at the shell may be acting on core magnetization components transverse to the field direction. 

\subsection{Loop asymmetries}
\begin{figure}[t]
\includegraphics[width=\columnwidth]{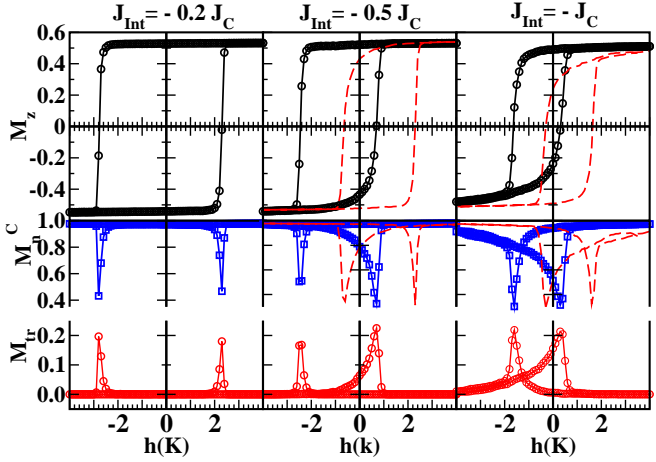}
\caption{\label{Fig_Asy}(Color online) Upper panels display the hysteresis loops  obtained after FC down in a field $h_\mathrm{FC}= 4$ (circles) and $h_\mathrm{FC}= -4$ (dashed lines) for three values of the exchange coupling constant $J_\mathrm{Int}$ at the core/shell interface. Lower panels show the average magnetization projection of the core spins along the field axis $m_\mathrm{n}^C$ (squares) and the hysteresis loops for the  component of the magnetization transverse to the field direction $M_\mathrm{tr}$ (circles). [Reprinted with permission from Ref. \onlinecite{Iglesias_jmmm07}, O. Iglesias et al. J. Magn. Magn. Mater. (in press, doi:10.1016/j.jmmm.2007.02.057) (2007). Copyright @Elsevier B. V.]}
\end{figure}

In addition, a clear asymmetry between the upper and lower loop branches developes when increasing the value of the interface coupling, as it is apparent when comparing the descreasing and increasing branches of the loops in the top panels of Fig. \ref{Fig_FCH}a or Fig. \ref{Fig_Asy}.
This feature can be more clearly seen by looking at the average absolute value of the magnetization projection along the field axis through the reversal process, $M_n^{\mathrm C}=\sum_i |\vec{S_i}\cdot\hat{z}|$, displayed in the middle panels of Fig. \ref{Fig_Asy} for the core spins. 
This quantity presents peaks centered around the coercive fields that indicate deviations of the core magnetization from the applied field direction. 
In the ZFC case, the peaks are centered at similar field values and they are quite narrow and almost symmetric around the minimum. However, for the FC loops, apart for the obvious shift of the peak positions, the decreasing branch peak is symmetric and narrow, while the increasing branch peak is deeper and asymmetric, enclosing bigger area under the loop curve. 

Asymmetric loops are usually found in different bilayered systems \cite{Engebretson_prb05,Camarero_prl05,Arenholz_apl05,Popova_epjb05,Olamit_prb06,Paul_prb06,Paul_jpcm06,Gredig_prb06} and are also evident in some core/shell nanoparticle systems. However, clear-cut experiments revealing the microscopic origin of this asymmetry have only been performed in the former case \cite{Fitzsimmons_prl00,Radu_prb03,Eisenmenger_prl05,Dekadjevi_prb06}. 
Most theories of EB for thin films, although considering the possibility of formation of domain walls during the magnetization reversal, are not able to account for origin of this asymmetry. Only in recent micromagentic simulations \cite{Kirschner_ieee03,Suess_prb03}, an asymmetry has been observed. Also MC simulations of the DS model for a single \cite{Beckmann_prl03} or twined anisotropy axes \cite{Beckmann_prb06} have shown that the observed asymmetries depend on the angle between the easy axis of the AFM and the applied magnetic field.  
More recently, hysteresis loops computed by MC simulations of a FM cylindrical dot in contact with an AFM based in a ferromagnetic domain wall model for the interfacial coupling, exhibited also an asymmetric profile \cite{MejiaLopez_prb05}. 
 
However, within the context of our model, in core/shell nanoparticles, the observed loop asymmetries arises solely by the competition between the interfacial exchange coupling and the aligning effect of the magnetic field due to the intricate geometry at the interface. 
\begin{figure*}[thbp]
\includegraphics[width=0.9\textwidth,angle= 0]{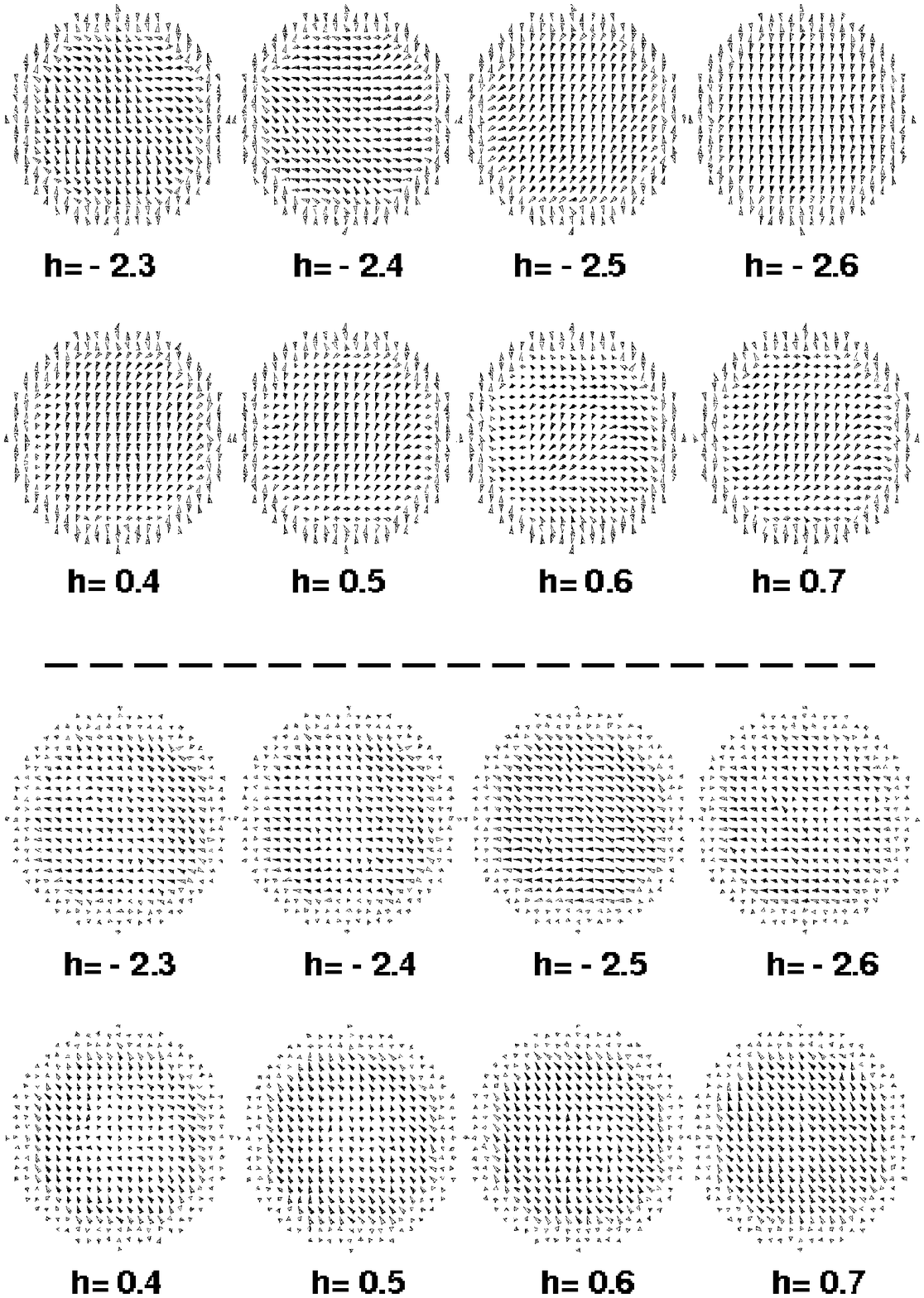}
\caption{\label{Fig_Configs}(Color online) Snapshots of the spin configurations of a midplane cross section of the particle parallel (upper sequence) and perpendicular (lower sequence) to the applied field direction taken at selected values of fields along the descending and ascending branches close to the coercive fields ($h_C^-$, $h_C^+$) for the case $J_{\mathrm {Int}}= -0.5 J_\mathrm{C}$ shown in Fig. 1a.
} 
\end{figure*}

\subsection{Reversal mechanisms}
These observations also indicate that the loop asymmetry reflects different reversal mechanisms in both branches of the hysteresis loops.
This can be corroborated by direct inspection of the spin configurations along the loops. In Fig. \ref{Fig_Configs}, several snapshots of a midplane cross section parallel (left panel) and perpendicular (right panel) to the applied field direction, taken around  the coercive fields $h_\mathrm{C}^\pm$ for $J_\mathrm{Int}= -0.5J_\mathrm{C}$ are shown. As it is evidenced by the upper sequence of snapshots, the reversal proceeds by quasi uniform rotation along the descending branch, while nucleation of reversed domains at the interface and their subsequent propagation through the core center is basically the reversal process along the ascending branch, as evidenced by the lower sequence in Fig. \ref{Fig_Configs}. Similar asymmetry between the loop branches has been also observed experimentally in bilayers \cite{Fitzsimmons_prl00,Radu_prb03,Eisenmenger_prl05} and, more recently, the relevance of nonuniform reversal modes to asymmetric magnetization reversal has been evidenced by measurements of hysteresis loops with varying angle of the cooling field in Ni/NiO polycrystalline system \cite{Dekadjevi_prb06}.
A detailed inspection of the configurations, also reveal the presence of spins at core/shell interface aligned perpendicular to the field direction for intermediate field values (see for example the snapshots for $h= -2.4, 0.6$ in Fig. \ref{Fig_Configs}). This observation corroborates the interpretation of recent results of small-angle neutron scattering experiments on Fe oxidized nanoparticles, in which the anisotropy of the obtained spectra was attributed to the existence of a net magnetic component aligned perpendicularly to the field direction \cite{Loffler_nanostr97,Ijiri_apl05}. Note that similar perpendicular couplings have been observed in thin film systems \cite{Moran_apl98,Ijiri_prl98}.

The microscopic origin for the different reversal mechanisms can be further clarified by looking at the behavior of the interface shell spins along the hysteresis loop (see Fig. \ref{Fig_FCH}b,c). While in the descending branch there is a considerable amount of unpinned spins that are able to reverse following the core reversal, in the ascending branch $M_\mathrm{Sh}^\mathrm{Int}$ remains constant (for $J_\mathrm{Int}<0$), an indication that spins at the shell interface remain pinned, hindering uniform rotation of the core but acting as a seed for the nucleation of reversed domains. 

\begin{figure}[tbp]
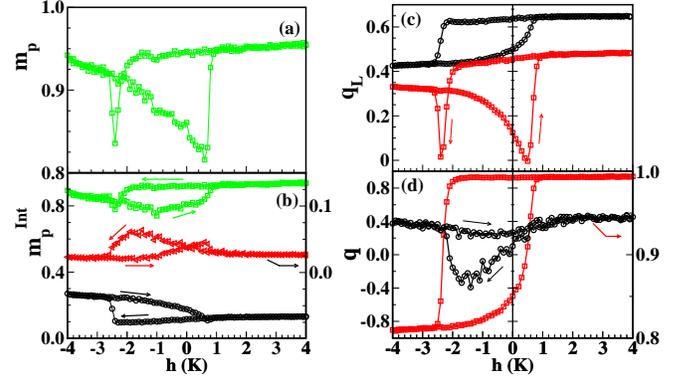

\includegraphics[width=0.49\columnwidth,angle= 0]{Mproj_new.eps}
\includegraphics[width=0.49\columnwidth]{FC+ZFC_Js-05_Ji-05_Overall2_vert.eps}
\caption{\label{Fig_qqL}(Color online) 
(Color online) Panel (a) shows the field dependence of the average spin projection of the core spins into the total magnetization direction $m_p^C$ (squares). In panel (b), only the contribution of all the interface spins (circles) has been taken into account, while the contributions of core and shell spins at the interface are shown in squares and triangles, respectively.
The field dependence of the link overlap $q_\mathrm L$ and overlap $q$ functions for the interfacial spins at the shell (circles) and at the core (squares) is shown in the panels at the right [(c) and (d), respectively], for $J_\mathrm{Int}= -0.5 J_\mathrm{C}$.
}
\end{figure}

The changes in the magnetic order at the core/shell interface and the presence of domain walls during reversal can be traced by monitoring the value of the average sum of the projection of the spin direction into the direction of the total magnetization vector along the hysteresis loops computed as 
\begin{eqnarray}
m_p(h) = \frac{1}{N} \sum_{i=1}^N \vec{S_i}(h)\cdot \vec{M_i}(h)\ .
\end{eqnarray}
This quantity should be close to $1$ if the magnetization reversal proceeds by uniform rotation of the spins, since in this case the spins remain parallel to the global magnetization direction. Deviations of $m_p(h)$ from $1$ indicate the formation of non-uniform structures during the reversal process. 
An example of the field variation of $m_p$ computed for all the core spins is shown in Fig. \ref{Fig_qqL}a, while in Fig. \ref{Fig_qqL}b we show $m_p(h)$ for the interfacial spins, where we have plotted separately the contribution of the core spins.  

During the decreasing field branch of the loop, $m_p$ remains quite close to $1$ for the core spins, except for moderate decrease down  for values of $h$ close to  $h_{\mathrm C}^-$ . The sharpness and symmetry of the peak around $h_{\mathrm C}^-$ confirms that the reversal proceeds by uniform rotation. In contrast, during the increasing field branch, an increasingly strong departure of $m_p$ from $1$ starting from negative field values can be clearly observed, reaching its maximum value also near $h_{\mathrm C}^+$. In this case, the observed peak asymmetry is indicative of the nucleation of the non-uniform domains observed in the snapshots of Fig. \ref{Fig_Configs}. 
These domains are formed at those points of the core interface with weaker values of the local exchange fields, as indicated by the more pronounced departure from $1$ of $m_p^{\mathrm {Int}}(h)$ (see Fig. \ref{Fig_qqL}b), than that corresponding to the total core magnetization (see Fig. \ref{Fig_qqL}a). 
The variation of $m_p^{\mathrm {Int}}$ for interface shell spins during the decreasing branch indicates the existence of a fraction of shell spins that reverse dragged by the spins at the core, while constancy of $m_p$ in the ascending branch is indicative of spins pinned during the core reversal.

The origin of the loop asymmetry can be further clarified by monitoring the values the so-called overlap $q(h)$ and link overlap $q_L(h)$ functions along the hysteresis loops, that are a generalization of similar quantities commonly used in the spin-glass literature \cite{Katzgraber_prb01,Katzgraber_prl02} and that are defined as  
\begin{eqnarray}
q(h) = \frac{1}{N} \sum_{i=1}^N \vec{S_i}(h_\mathrm{FC})\cdot \vec{S_i}(h)\nonumber\\
q_L(h) =  \sum_{\langle ij \rangle} \frac{1}{N_l}\ \vec{S_i}(h_\mathrm{FC})\cdot \vec{S_j}(h_\mathrm{FC})\ \vec{S_i}(h)\cdot\vec{S_j}(h)\ ,	
\end{eqnarray}
where in $q_L(h)$ the summation is over nearest neighbors and $N_l$ is a normalization factor that counts the number of bonds. 

An example of the field dependence of these overlaps, computed only for the interfacial spins, is shown in Figs. \ref{Fig_qqL}c, d, where we have separated the contribution of the shell and core spins.
A departure of $q_L$ from unity is known to be proportional to the surface of reversed domains formed at field $h$ and, therefore, $q_L$ is sensible to the existence of non-uniform structures.
The sharp decrease of $q_L$ for core spins and the symmetry of the peak around the negative coercive field indicates uniform reversal. However, the progressive reduction of $q_L$ along the ascending branch and the asymmetry of the peak around the positive coercive field indicates the formation of reversed nuclei at the particle core that sweep the particle during reversal.

The function $q(h)$ measures differences of the spin configuration at field $h$ with respect to the one attained after FC. Therefore, the decrease of $q$ for the interface shell spins when reducing the magnetic field indicates the existence of a fraction of shell spins that reverse dragged by core spins, while the constancy of $q$ in the ascending branch reveals the existence of spins pinned during core reversal. 

\subsection{Vertical loop shifts}
\begin{figure}[tbp]
\includegraphics[height=0.99\columnwidth,angle= -90]{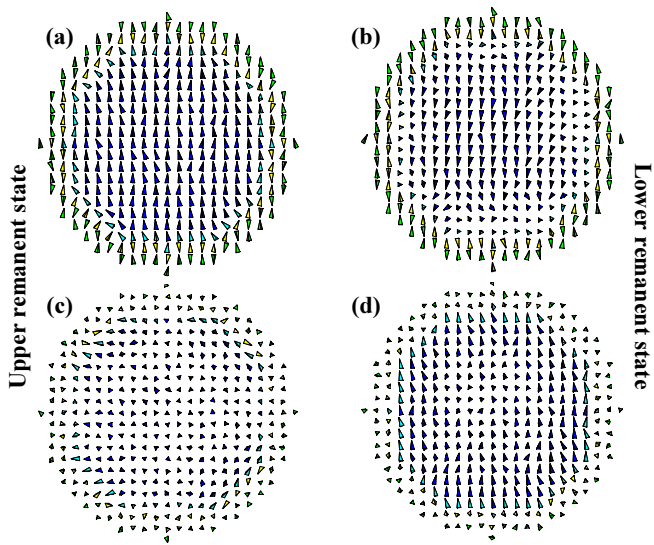}
\caption{\label{Fig_ConfigsRem}(Color online) Snapshots of the remanent spin configurations of the upper (a,c panels) and lower (b,d panels) branches of the hystersis loops showing midplane cross sections of the particle parallel (a,b) and perpendicular to the z axis (c, d) for the case $J_{\mathrm {Int}}= - J_\mathrm{C}$ shown in Fig. 1. Dark (light) blue cones represent core (core/interface) spins while green (yellow) ones are for spins at the shell (shell/interface).
}
\end{figure}

Clearly correlated to the observation of EB and the loop asymmetry, the loops experience a shift along the vertical ($M_z$) axis which increases with $J_\mathrm{Int}$, as reflected in Fig. \ref{Fig_Asy} by the difference of the $M_z$ values in the high field region of the two loop branches or at the remanence points. 
The field dependence of magnetization component transverse to the field direction, $M_\mathrm{tr}$ (circles in the lower panels of Fig. \ref{Fig_Asy}), indicates that $M_\mathrm{tr}$ attains values for the descending loop branch that are higher than in the ascending branch. Moreover, the $M_\mathrm{tr}$ values around the peaks increase with increasing $J_\mathrm{Int}$. 
Snapshots of the spin configurations at the remanence points of the hysteresis loops are displayed in Fig. \ref{Fig_ConfigsRem}. They show the existence of a higher amount of core spins with transverse orientation near the interface at the lower branch (Fig. \ref{Fig_ConfigsRem}b, d) than at the upper branch. This is in agreement with the results of some experiments in oxidized particles where this vertical shift was also observed \cite{Zheng_prb04,DelBianco_prb04,Tracy_prb05,Passamani_jmmm06,Ceylan_jap06,Tracy_prb06} and with the observation  of transverse magnetization components during reversal revealed unambiguously by magneto-optical Kerr effect in bilayers \cite{Tillmanns_apl06}. Our simulation results above, indicate that the microscopic origin of the vertical shift resides in the different reversal mechanisms on the two loop branches due to the existence of uncompensated pinned moments at the core/shell interface that facilitate the nucleation of non-uniform magnetic structures during the ascending field branch of the loops. The recent observation that the vertical shift may be attributed to the existence defect moments \cite{Tracy_prb06} will be checked within the scope of our model by removing some spins at the interface or at the core of the particle, this work is in progress. 

\subsection{Temperature dependence of h$_{eb}$}

The thermal dependence of $h_\mathrm{eb}$ and $h_\mathrm{C}$ can also be studied by the MC method. Using a model for an oxidized nanoparticle similar to ours, Trohidou and coworkers \cite{Zianni_jpcm98} first computed the thermal dependence of the coercive field founding that, compared to the non-oxidized particles, there was an increase of $h_\mathrm{C}$ in all the temperature range and also a reversal in the size dependence of the coercivity. They also found a steeper temperature dependence of $h_\mathrm{C}$ when the  interface anisotropy is enhanced. More recently, they have also computed thermal dependences of $h_\mathrm{eb}$ for several particle sizes \cite{Eftaxias_pssc04,Eftaxias_prb05,Trohidou_jmmmun}, finding a stronger temperature dependence for the bias field than for the coercive field. 
MC simulations of the DS model for bilayers \cite{Nowak_prb02} have also found a linear decrease of $h_\mathrm{eb}$ vanishing below $T_N$, in excellent agreement with experimetal results \cite{Keller_prb02}.

\begin{figure}[tbp]
\includegraphics[width=\columnwidth]{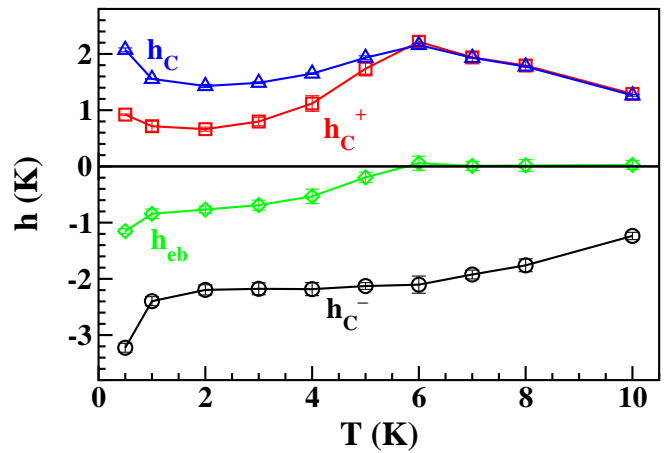}
\caption{\label{Fig_temp}(Color online) Thermal dependence of the coercive fields $h_\mathrm{C}^-$, $h_\mathrm{C}^+$, $h_\mathrm{C}$ and the exchange bias field $h_\mathrm{eb}$ for a particle with $J_\mathrm{Int}= -0.5 J_\mathrm{C}$.
}
\end{figure}

The results of our finite temperature simulations for the particle with $J_\mathrm{Int}= -0.5 J_\mathrm{C}$ are displayed in Fig. \ref{Fig_temp}, where the variation of $h_\mathrm{C}^-$, $h_\mathrm{C}^+$, $h_\mathrm{C}$ and $h_\mathrm{eb}$ with the temperature at the end of the FC process are displayed. Let us notice first the different dependencies of $h_\mathrm{C}^\pm$ on $T$. 
Starting from the lowest temperature, both quantities first decrease up to $2$ K aprox. However, after further increase in $T$, whereas $h_\mathrm{C}^-$ is stable up to $T_B= 6$ K, $h_\mathrm{C}^-$ increases, reaching a maximum at the same $T$.
As a consequence, we find that $h_\mathrm{eb}$ vanishes at 6 K, while $h_\mathrm{C}$ presents a maximum at the same temperature. This seems to agree with Trohidou's results \cite{Eftaxias_prb05} for some of their particle sizes. This enhancement of $h_\mathrm{C}$ at the blocking temperature $T_B$ where $h_\mathrm{eb}$ vanishes has also been reported for bilayered systems, but, to our knowledge, not for particles. 

\subsection{Other studies}
The particle size and shell thickness dependence of EB has been studied by Trohidou and coworkers \cite{Eftaxias_prb05}. These authors argue that the observation of EB depends mainly on the structure of the interface and not on its size, also in agreement with our findings. They have found that a reduction of the core size for a given particle size enhances $h_\mathrm{eb}$ and reduces $h_\mathrm{C}$. 
The same group \cite{Fiorani_prb06,Trohidou_jmmmun} has recently performed simulations of a core/shell nanoparticle with random anisotropy directions in a FIM shell which reproduce the experimentally found training effects in Fe oxidize particles \cite{DelBianco_prb04,DelBianco_jmmm05,DelBianco_jmmmun} and also the aging of the remanent magnetization, $h_\mathrm{C}$ and $h_\mathrm{eb}$. This last quantity was found to increse with the time during which the cooling field has been applied. 
The existence of training effects in layered systems has also been shown to be in agreement with experiments in MC simulations within the context of the DS model \cite{Miltenyi_prl00,Nowak_prb02}. The role of imperfect interfaces in establishing the EB has also been studied in this model \cite{Spray_jpd06}. 
Let us also mention the works by Usov at al. \cite{Usov_jmmm05,Usov_jmmm06,Usov_jmmmun}, in which the authors compute the magnetic states and hysteresis loops of composite nanoparticles and bilayers using a quantum mechanical Hartree-Fock approximation. 
Also using a quantum mechanical approach, Mata et al. \cite{Mata_prb06} have suggested that quantum zero temperature fluctuations of surface spins near an AFM surface induce dipole fields that may account for the observed exchange anisotropies. 

A microscopic model for interface roughness in bilayers was proposed by Almeida and Rezende \cite{Almeida_prb02}, who computed the hysteresis loops for Ising spins in a mean-field approximation. Apart from the loop shift and enhanced coercivity, they showed that the sign of the exchange bias field changes as the initial temperature of the FC process is lowered and as the cooling field is varied, in agreement with experimental reports \cite{Leighton_prl00,Nogues_prl96}. In a model of bilayers based on a generalization of MB model that included biquadratic exchange and that accounted for the granular structure of the AFM, Hu et al. \cite{Hu_jap02,Hu_jap03,Hu_epjb04} computed the thermal dependence of $h_\mathrm{C}$ and $h_\mathrm{eb}$, in agreement with some experimental results.  

MC simulations by Lederman et al. \cite{Lederman_prb04} for Fe/FeF$_2$ demonstrated that EB is generated when the AF sublattices have an unequal exchange coupling with the FM and that perpendicular order between the FM and AFM is possible for large interface exchange coupling, in agreement with previous theories \cite{Kiwi_epl99,Kiwi_apl99}. In similar MC simulationsy, Billoni et al. \cite{Billoni_phaB06} have studied the influence of the value of exchange coupling constant at the interface on $H_{eb}$ at different temperatures.
The effect of interfacial coupling on the magnetic ordering of models of coupled bilayers was studied using MC simulations by Tsai {\slshape et al.} \cite{Tsai_jap02,Tsai_jap03} and Alonso {\slshape et al.} \cite{Alonso_prb06} , by Finazzi \cite{Finazzi_prb04} in a micromagnetic approach. 

Within the context of a random field Ising model, Illa et al. \cite{Illa_prb02,Illa_jmmm04} performed MC calculations of bilayers where the existence of EB was related to the fraction of enhanced broken links and was shown to be due to minor loop effects. The same approach has been used in a model that includes a partially covering of the FM/AFM interface by a non-magnetic spacer, showing its influence in perpendicular EB \cite{Blachowicz_msp05,Blachowicz_cejp05,Blachowicz_cejp06}. Also based on the same model, Meilikhov and Farzetdinova \cite{Meilikhov_jetp05} presented a mean-field approach that allows analytical solutions. MC simulations of the related random anisotropy Ising model by Negulescu et al. \cite{Negulescu_joam04} showed also EB effects due to the roughness of the interface. 

First principle studies specifically addressing the origin of the EB effect are scarce. However, interesting calculations of Co/FeMn bilayers by Nakamura et al. \cite{Nakamura_prb04} using FLAPW method to incorporate noncollinear magnetic structures have demonstrated from first principles that an out-of-plane magnetic anisotropy is induced at the Co/FeMn interface, in accordance with experimental reports.

\section{Concluding remarks}

We have reviewed the main phenomenology associated to EB in core/shell nanoparticles and presented details of our simulations of a model for these systems which explicitly takes into account the microscopic parameters characterizing the core/shell interface. The results of the simulations are able to account for some of the experimental observations. 
The obtained hysteresis loops after FC present shifts along the field axis which are directly related to the existence of a fraction of uncompensated spins at the shell interface that remain pinned during field cycling.
The results of the simulations have revealed asymmetries in the hysteresis loops which, by detailed analysis of the microscopic magnetic configurations, have been linked to the occurrence of different reversal mechanisms in the two loop branches. The existence of  interfacial groups of spins aligned transverse to the field direction and the above mentioned difference in the reversal mechanisms is also responsible for the vertical shift of the loops. Moreover, we have been able to establish a quantitative connection between the macroscopic magnitude of the EB fields and the microscopic value of the net magnetization of the interfacial shell spins.

In order to account for the effects that other characteristic features of real nanoparticle systems may have on the experimentally observed phenomenology, further ingredients will have to be considered in simulations of microscopic models. 
Among them, let us mention the intrinsic surface spin disorder and surface anisotropy, the distribution in particle sizes and randomness of the anisotropy directions and the existence of dipolar interparticle interactions in self-assembled or agglomerated particle assemblies.  
Finally, we hoped that the possibility to perform ab initio calculations of nanoscale clusters form first principles will lead to more realistic inputs for the microsocopic parameters needed for MC and micromagnetics simulations, allowing a multiscale approach that will shed new light into the microscopic origin of the peculiar magnetic properties of nanoparticles.

\begin{acknowledgments}
We acknowledge CESCA and CEPBA under coordination of C$^4$ for computer facilities. This work has been supported by the Spanish MEyC through the MAT2006-03999, NAN2004-08805-CO4-01/02 projects and the Generalitat de Catalunya through the 2005SGR00969 DURSI project. We are indebted to J. Nogués for careful reading of the manuscript and useful comments and suggestions.
\end{acknowledgments}

\end{document}